\documentclass{article}
\usepackage{spconf,amsmath,graphicx}
\usepackage{booktabs}
\usepackage{xcolor}
\usepackage{float}

\usepackage{amsmath,amsfonts,bm}

\def\eqref#1{equation~\ref{#1}}

\def\1{\bm{1}}

\def\vh{{\bm{h}}}

\def\vx{{\bm{x}}}

\DeclareMathAlphabet{\mathsfit}{\encodingdefault}{\sfdefault}{m}{sl}
\SetMathAlphabet{\mathsfit}{bold}{\encodingdefault}{\sfdefault}{bx}{n}

\title{Do We Still Need Automatic Speech Recognition\\for Spoken Language Understanding?}
\name{\parbox{\textwidth}{\centering Lasse Borgholt$^{1, 3}$, 
      Jakob D. Havtorn$^{2, 3}$,
      Mostafa Abdou$^{1}$,
      Joakim Edin$^{3}$,\\
      Lars Maaløe$^{2, 3}$, 
      Anders Søgaard$^{1}$, and
      Christian Igel$^{1}$}}
\address{
  $^1$Department of Computer Science, University of Copenhagen, Denmark\\
  $^2$Department of Applied Mathematics and Computer Science, Technical University of Denmark\\
  $^3$Corti, Copenhagen, Denmark\\
  borgholt@di.ku.dk}

\begin{document}
%
\maketitle
\begin{abstract}
Spoken language understanding (SLU) tasks are usually solved by first transcribing an utterance with automatic speech recognition (ASR) and then feeding the output to a text-based model. Recent advances in self-supervised representation learning for speech data have focused on improving the ASR component. We investigate whether representation learning for speech has matured enough to replace ASR in SLU. We compare learned speech features from wav2vec 2.0, state-of-the-art ASR transcripts, and the ground truth text as input for a novel speech-based named entity recognition task, a cardiac arrest detection task on real-world emergency calls and two existing SLU benchmarks. We show that learned speech features are superior to ASR transcripts on three classification tasks. For machine translation, ASR transcripts are still the better choice. We highlight the intrinsic robustness of wav2vec 2.0 representations to out-of-vocabulary words as key to better performance.



\end{abstract}
\section{Introduction}
\label{sec:intro}


Pre-training large transformers with self-supervised learning (SSL) has dramatically advanced automatic speech recognition (ASR) \cite{baevski2020wav2vec} and shown promise for taking spoken language understanding (SLU) to the next level \cite{lai2021semi, yang2021superb}.  However, it remains an open question whether better ASR is the key to improving SLU or if the speech representations learned with SSL offer a better alternative as input to downstream models. Labeled data for ASR training can be difficult to obtain, so if self-supervised representations are a competitive alternative, this has the potential to democratize SLU.


Work on speech representation learning has primarily focused on ASR \cite{baevski2020wav2vec} and other speech-specific tasks, such as emotion recognition \cite{pascual2019learning}, speaker identification \cite{liu2020mockingjay} and phoneme classification \cite{baevski2021unsupervised}. Tasks that require high-level semantics have been studied less \cite{lai2021semi, yang2021superb}. Accordingly, the number of SLU tasks is limited, and many text-based natural language understanding tasks cannot be directly translated to the speech domain due to the difficulty of obtaining word segmentation. For this reason, existing tasks only contain little data \cite{lugosch2019speech} or make use of synthetic speech \cite{yang2021superb}.

\begin{figure}[!t]
  \centering
  \vspace{7mm}
  \includegraphics[width=\linewidth]{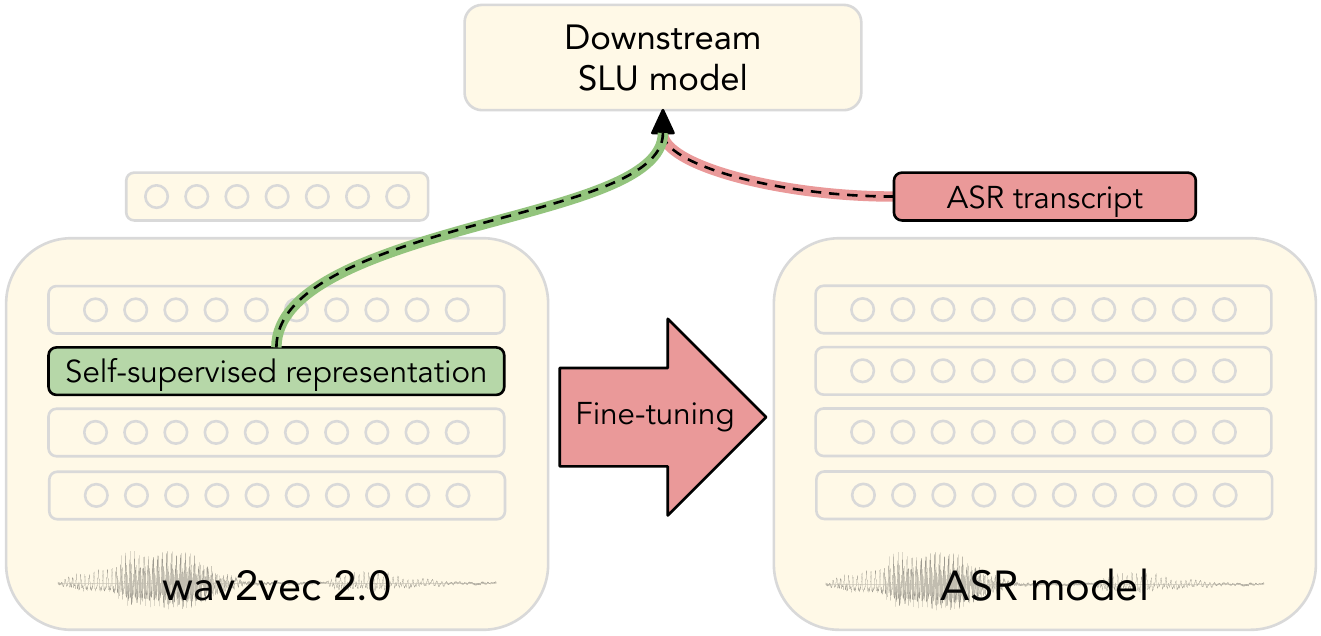}
  \caption{Self-supervised models, such as wav2vec 2.0, yield state-of-the-art results when fine-tuned for automatic speech recognition. However, we show this step is redundant for many downstream spoken language understanding tasks where self-supervised representations can be used as input.}
  \label{fig:exp_setup}
  \vspace{5mm}
\end{figure}


In this work, we compare speech features from the wav2vec~2.0 model, state-of-the-art ASR transcripts and ground truth text as input for four SLU tasks that all require knowledge about high-level semantic concepts. While the quality of an ASR model can always be debated, the ground truth text defines an upper bound on ASR performance and serves as a baseline. We consider existing intent classification (IC) and machine translation (MT) tasks and present a novel speech-based named entity recognition (NER) task. Finally, we use a proprietary dataset of 911-calls to define a noisy real-world task of cardiac arrest detection (CAD). Our contributions are as follows:

\begin{table*}[t]
\begin{center}
\begin{tabular}{ l | c c c | c c c } 
\toprule

 & \multicolumn{3}{c|}{\textbf{Number of examples}} & \bf Duration & \bf Median & \bf WER \\
\textbf{Task - Dataset} & train & valid & test & [h] & [s] & [\%] \\

\midrule

Named entity recognition - LibriSpeech & 281,241 & 5,567 & 5,559 & 982 & 16.7 & 2.1 / 4.5 \\

Cardiac arrest detection - Seattle Fire Department & 2,086 & 260 & 260 & 146 & 171.5 & 35.1 \\

Intent classification - Fluent Speech Commands & 23,132 & 3,118 & 3,793 & 19 & 1.8 & 23.2 \\

Machine translation - CoVoST 2 (En-De) & 289,165 & 15,517 & 15,524 & 478 & 5.4 & 16.6 \\


\bottomrule
\end{tabular}
\end{center}
\vspace{-0.4cm}
\caption{Basic dataset statistics for the SLU tasks. \textit{Number of examples} in the subsets. \textit{Duration} in hours for all subsets. \textit{Median} example length in seconds. Word error rate (WER) on the validation set for the ASR models presented in \ref{sec:speech_asr}.} 
\label{tab:tasks}
\end{table*}

\begin{enumerate}
\setlength\itemsep{0mm}
\item We present the first systematic comparison between text and speech features as input to a broad range of SLU tasks since the recent surge in SSL for speech.
\item We show that wav2vec 2.0 speech representations are better than ASR transcripts for downstream IC, NER and CAD tasks when labeled data is limited and competitive when data is abundant. For MT, the speech representations fall short of text. 
\item We introduce a speech-based NER task derived from the most widely used dataset for research in ASR and SSL for speech, LibriSpeech.
\end{enumerate}

\section{Tasks}
\label{sec:tasks}

We start by presenting the four tasks, all based on natural speech. Dataset statistics are found in table \ref{tab:tasks}. 

\subsection{Named entity recognition: LibriSpeech}
\label{sec:ner-task}



LibriSpeech \cite{panayotov2015librispeech} is derived from audiobooks part of the LibriVox project\footnote{https://librivox.org}. Training data for wav2vec 2.0 consist of 60K hours of speech from LibriVox, while the open-source ASR models used in this work are trained on LibriSpeech unless stated otherwise. Defining a downstream task on data from the same domain used to train the SSL model and ASR model corresponds to a common scenario where training data for the different modeling steps overlap. LibriSpeech comes with multiple standardized training subsets \cite{kahn2020libri} allowing us to study how the downstream model is affected by varying the amount of training data. Finally, LibriSpeech contains two validation and test subsets, \textit{clean} and \textit{other}, which offer insight into the importance of recording quality.

We provide silver label named entity tags for LibriSpeech. The labels were obtained by using an off-the-shelf Electra language model\footnote{https://huggingface.co/dbmdz} fine-tuned on the CoNLL-2003 NER task \cite{clark2019electra, wolf2019huggingface} and applying the model to the ground truth text. We manually reviewed model-induced labels for the validation set to get a sense of data quality. For 1,000 randomly selected examples, human-model agreement is high, with a Krippendorff's alpha of 0.98.

The task is to predict whether a named entity is contained in the input example. In contrast to classic text-based NER, where each word is tagged with a label, we considered this a binary sequence-based classification task to keep the model setup for text and speech features as similar as possible.

Conveniently, we find that the dataset is balanced such that approximately 50\% of the examples in the training subsets contain a named entity. For validation and test, the fraction is around 30\%. We make the dataset available with more details\footnote{https://github.com/borgholt/ner-librispeech}, such as entity type (i.e., \textit{person}, \textit{location}, \textit{organization} or \textit{miscellaneous}) and entity alignment obtained from \cite{lugosch2019speech} which used the Montreal Forced Aligner\footnote{https://montreal-forced-aligner.readthedocs.io}.


\begin{table*}[t]
\begin{center}
\begin{tabular}{ l  l | c c c c c c } 
\toprule
 
 & & \multicolumn{2}{c}{\textbf{10 hours}} & \multicolumn{2}{c}{\textbf{100 hours}} & \multicolumn{2}{c}{\textbf{960 hours}}\\
 
\textbf{Input} & \textbf{Source} & clean & other & clean & other & clean & other\\

\midrule
Characters & GT & 57.9 & 60.6 & 79.2 & 80.5 & 91.0 & 92.0 \\
Word & GT & 58.9 & 58.7 & 75.2 & 76.4 & 89.2 & 91.0 \\
\midrule
Characters & ASR & 56.4 & 59.0 & 78.0 & 79.9 & \textbf{90.3} & \textbf{88.2} \\
Word & ASR & 56.0 & 58.7 & 73.3 & 75.7 & 88.1 & 88.2 \\
Wav2vec 2.0 & SSL & \textbf{83.0} & \textbf{81.7} & \textbf{88.5} & \textbf{83.6} & 90.0 & 86.3 \\

\bottomrule
\end{tabular}
\end{center}
\vspace{-0.4cm}
\caption{Named entity recognition results on the LibriSpeech test sets. All results are given in F1-scores.} 
\label{tab:ner}
\end{table*}

\subsection{Cardiac arrest detection: Seattle Fire Department}
\label{sec:ca-task}

From a proprietary dataset of 911 emergency calls provided by Seattle Fire Department, WA, USA, we constructed a binary sequence classification task where the objective is to predict whether the caller describes an out-of-hospital cardiac arrest (OHCA) or not. The original dataset contains 1303 OHCA calls and many more not-OHCA calls. We did a random 80-10-10 split of the OHCA calls and sampled a not-OHCA call of similar length to each of the OHCA calls to keep the dataset balanced in terms of target distribution and hours of speech per class. We did not have ground truth text available for this task but report word error rate on a separate subset in table \ref{tab:tasks}.

\subsection{Intent classification: Fluent Speech Commands}
\label{sec:intent-task}

The Fluent Speech Commands (FSC) dataset \cite{lugosch2019speech} consists of 248 unique read-speech commands from 97 speakers instructing a hypothetical intelligent home device to perform an action (e.g., "Turn the lights on in the kitchen"). Recording of the commands was crowd-sourced, resulting in a varied selection of English speakers from the US and Canada. The task was originally phrased as a multi-slot task with three slots: \textit{action}, \textit{object} and \textit{location}. However, due to the small number of slot classes, the task is commonly rephrased as a simple classification task with 31 unique classes.

\subsection{Machine translation: CoVoST 2}
\label{sec:mt-task}

CoVoST 2 is a multilingual speech-to-text translation dataset \cite{wang2020covost} derived from the Common Voice speech corpus \cite{commonvoice:2020}. Translations were made by professional translators and the corresponding speech recordings were crowd-sourced. We focused on the English-to-German task using the so-called CoVoST training set and the Common Voice test and validation sets as in the original work.

\section{Experiments}
\label{sec:exps}

\subsection{Task-specific models}
\label{sec:Model}

We are interested in comparing the information content of the \textit{input representations}, not the models, so we chose a minimalist architecture. All models take as input a sequence of vectors $\displaystyle \vx_{1:T} =  \vx_1, \vx_2, \dots, \vx_T$ where $\vx_t\in\mathbb{R}^{K}$ and share a similar encoder. A fully connected layer without activation maps each $\vx_t$ to a $D$-dimensional linear subspace. This linear mapping is the only source of variation in terms of model parameterization between the input representations as it depends on the input dimensionality $K$; 1024 for wav2vec 2.0 representations, 29 for character-level text and 1,296 to 41,341 for word-level text. The linearly projected features are fed to a bidirectional LSTM with a $D$-dimensional recurrent state.

Hereafter, each task requires a different architecture. For the binary NER and CAD tasks, the LSTM output $\vh_{1:T}$ is max-pooled and fed into a single neuron with a sigmoid activation to parameterize a Bernoulli distribution. Similarly, for the IC task, the LSTM output is pooled and mapped to a 31-dimensional vector with softmax normalization to parameterize a categorical distribution. For the MT task, we used an LSTM-based autoregressive decoder with scaled dot-product attention \cite{vaswani2017attention}. We used a vocabulary of 10K subword units\footnote{https://github.com/google/sentencepiece} for the target language.

\subsection{Speech representations and ASR models}
\label{sec:speech_asr}
The wav2vec 2.0 models use contrastive self-supervised learning and are fine-tuned for ASR with a connectionist temporal classification loss. See \cite{baevski2020wav2vec} for more details. We considered two SSL-ASR model pairs downloaded from the \textsc{fairseq} sequence modeling toolkit\footnote{https://github.com/pytorch/fairseq}. For the first pair, the self-supervised wav2vec2.0~model has been trained on 60K hours of speech from LibriLight and fine-tuned on 960 hours from LibriSpeech \cite{baevski2020wav2vec}. The second pair, which is more robust, adds 3000 hours of conversational and crowd-sourced speech from the Fisher, Switchboard and CommonVoice corpora to the self-supervised training, while the ASR model was fine-tuned using the 300 hours from Switchboard \cite{hsu2021robust}. All models use the same architecture. We tested the two ASR models on the validation set for each task and chose the model pair corresponding to the lowest word error rate. For the IC and CAD tasks, the robust ASR model was better. 


As shown in \cite{baevski2021unsupervised, borgholt2021scaling}, the top layers of wav2vec~2.0 are a poor choice of input to phoneme classification and ASR. We ran a small initial experiment with limited training data to determine which output from the 24 transformer layers in the wav2vec 2.0 architecture to use as input to the downstream tasks. We found that layer 15 yielded the best results. 
This layer has also been found to provide the best results for phoneme classification \cite{baevski2021unsupervised}, and layers 13 through 16 have been shown to contain the highest level of correlation with text-based word embeddings \cite{pasad2021layer}.


\subsection{Training}
\label{sec:training}
All models were trained by minimizing cross-entropy and use $D = 256$. In the very low-resource settings, we also tested smaller dimensionalities to reduce overfitting, but this did not improve results. We used the Adam optimizer \cite{kingma2014adam} with a fixed learning rate of $3 \cdot 10^{-4}$ for the first half of training before annealing it to $5 \cdot 10^{-5}$ during the second half. Batch size and validation frequency were tuned for each task on the ASR character-level. We ensured that the number of training steps was large enough to reach convergence for all experiments.



\section{Results}
\label{sec:results}

Results for each of the four tasks are presented below. We report the metric commonly used in previous work for the existing tasks. \textit{GT} refers to \textit{ground truth} in tables 2 and 3.

\subsection{Named entity recognition: LibriSpeech}
\label{sec:ner-res}

The wav2vec 2.0 representations showed impressive performance on the 10-hour subset, as seen in table \ref{tab:ner}, where text-based models were only slightly better than a random baseline. Even with 100 hours of labeled data, they were superior. The gap closed at 960 hours. In general, models trained on ground truth text performed better on the \textit{other} subset, whereas speech-based models always performed best on the \textit{clean} subset, highlighting the speech features' sensitivity to noisy conditions. Although the ASR transcripts are also affected by noise, they gave more robust results, as these models performed better on the \textit{other} subset in all but one case. 

On examples that exclusively contain named entities that are out-of-vocabulary, wav2vec 2.0 representations gave an error rate of 23\% when trained on 100 hours. ASR transcripts gave a substantially higher error rate of 36\%. This underscores the large amount of data needed for robust out-of-vocabulary named entity recognition.

\subsection{Cardiac arrest detection: Seattle Fire Department}
\label{sec:ca-res}

Considering the observation that ASR transcripts are more noise-robust than wav2vec 2.0 representations, we might expect them to fare better on noisy 911-calls. However, as seen in table \ref{tab:cad_ic_mt}, the wav2vec 2.0 representations still yielded better results. Unlike the NER task, it is possible that speech-based features, such as emotion and rate of speech, might prove useful for this task.


\subsection{Intent classification: Fluent Speech Commands}
\label{sec:intent-res}

As mentioned in the task description, every speaker in this dataset read the same 248 commands. As a result, training, validation and test subsets contain the same 248 identical examples when we consider ground truth text, which leads to an accuracy of 100\% as seen in table \ref{tab:cad_ic_mt}. While the task is generally considered to require semantic understanding \cite{yang2021superb}, which is also why we include it here, 
it can be solved to perfection by a many-to-one sentence recognizer (i.e., different sentences map to the same intent). 
The wav2vec 2.0 representations were slightly better than the ASR transcripts and very close to the more complex state-of-the-art ASR-based system from \cite{qian2021speech} which reached 99.7\% accuracy.

\subsection{Machine translation: CoVoST 2}
\label{sec:mt-res}

Unsurprisingly, we find that our simple ASR-based MT system was a lot worse than ground truth text. The wav2vec 2.0 representations were even worse. These results are not surprising considering the generally large gap between speech and text-based approaches \cite{wang2020covost}. We hypothesize that the lack of simple morphological features, like word boundaries, is a challenge to overcome for a shallow model trained on speech-based representations. To test this hypothesis, we trained the model on the ASR character-level transcripts without white-spaces (e.g., \textsc{how are you} $\rightarrow$ \textsc{howareyou}) which resulted in a notable drop from BLEU 11.5 to 9.7, but not enough to explain the gap between the two representations.

\begin{table}[t]
\begin{center}
\begin{tabular}{ l  l | c c c} 
\toprule
 
 & & \textbf{CAD} & \textbf{IC} & \textbf{MT}\\
\textbf{Input} & \textbf{Source} & F1-score & Accuracy & BLEU\\

\midrule
Characters & GT & \textsc{n/a} & 100.0 & 15.5 \\
Word & GT & \textsc{n/a} & 100.0 & 14.0 \\
\midrule
Characters & ASR & 84.1 & 99.4 & \textbf{11.5} \\
Word & ASR & 82.5 & 98.7 & 10.9 \\
Wav2vec 2.0 & SSL & \textbf{84.7} & \textbf{99.6} & 6.1 \\

\bottomrule
\end{tabular}
\end{center}
\vspace{-0.4cm}
\caption{Results for cardiac arrest detection, intent classification and machine translation.} 
\label{tab:cad_ic_mt}
\end{table}

\section{Discussion}
\label{sec:discussion}

This work should not be seen as a quest to remove ASR from the SLU pipeline. Automatically generated transcripts offer an important layer of interpretability in modern speech applications. Furthermore, we did not explore how text-based language models can be modified to handle error-prone transcripts, which is a promising direction for SLU \cite{lai2021semi}. However, this work is highly relevant when large quantities of unlabeled speech data can be easily obtained, but no or limited text data is readily available -- such as in an emergency call center.

Our work suggests that ASR fine-tuning can be avoided for downstream SLU tasks. Interestingly, it was recently found that \textit{word meaning} is shifted towards the output layer of the model when wav2vec 2.0 is fine-tuned for ASR \cite{pasad2021layer}. Our work highlights the feasibility of extracting this knowledge directly from the pre-trained model.








\section{Conclusion}
\label{sec:conclusion}

We compared self-supervised speech features from wav2vec 2.0 with automatic speech recognition transcripts and ground truth text as input to a simple model on four spoken language understanding tasks. Interestingly, it turned out that wav2vec 2.0 representations yielded better performance than speech recognition transcripts with up to 100 hours of training data for cardiac arrest detection, named entity recognition and intent classification. Only when 960 hours of labeled training data was available, the speech recognition-based approach yielded a slight improvement on the named entity recognition task. For machine translation, the wav2vec 2.0 representations were inferior to the text-based features. Our results on the classification tasks have implications for how to tackle spoken language understanding tasks with limited training data demonstrating that the traditional automatic speech recognition step can be bypassed.

\vfill
\pagebreak

\bibliographystyle{IEEEbib}
\bibliography{refs}

\end{document}